\documentstyle[prl,aps,multicol,epsfig]{revtex}
\begin{document}
\draft
\title{Dynamical Mean Field Theory for Perovskites}
\author{P. Lombardo,$^a$ J. Schmalian,$^b$ M. Avignon,$^a$ 
and K.-H. Bennemann$^b$}
\address{ $^a$LEPES-CNRS, BP 166, 38042 Grenoble Cedex 9, 
France \\
$^b$ Institut f\"ur Theoretische Physik, Freie Universit\"at Berlin, 
Arnimallee 14, 14195 Berlin, Germany}
\date{\today}
\maketitle
\begin{abstract}
\leftskip 54.8pt
\rightskip 54.8pt
Using the Hubbard Hamiltonian for transition metal-3d and oxygen-2p 
states with perovskite
geometry, we present a dynamical mean field theory which becomes exact
in the limit of large coordination numbers or equivalently large 
spatial dimensions $D$.
The theory is based on  a new description   of
these systems  for large  $D$  using  a selective treatment of different
hopping processes  which can not be generated by a unique scaling 
of the hopping element.
The model is solved  using a perturbational approach and an extended 
non-crossing approximation.
We discuss the breakdown of the perturbation theory near half filling, the 
origin of the various 3d and 2p bands, the doping dependence
of its spectral weight, and 
the evolution of quasi particles at the Fermi-level upon doping, 
leading to interesting insight 
into the dynamical character of the charge carriers near  the metal insulator 
instability of transition metal oxide systems, three 
dimensional perovskites and other strongly correlated transition metal oxides.
\end{abstract}
%\newpage
\begin{multicols}{2}   
%
%
%\narrowtext
%
\section{Introduction}
\label{sec:Introduction}
The electronic structure of the  strongly correlated copper oxide 
superconductors, other transition metal   based perovskites, 
and of  transition metal oxides   like NiO, FeO or MnO  has been an enduring
problem in the last  years~\cite{BM86,FTA91,ATT93,H95,SD95,SBM95}, 
still containing numerous unresolved questions.
Among them are the  nature of the insulating state, the
evolution and doping dependence of coherent quasi particles near the
Fermi energy in a doped Mott Hubbard or charge transfer insulator~\cite{ZSA85} and the
transfer of spectral weight from high to low energy scales~\cite{EMS91}.
Experimentally, interesting variations of the spectroscopic and thermodynamic
 properties of these materials occur
upon electron and hole doping~\cite{CRH91,TTO93,KKG89,RSH95}.
Recently, Tokura {\em et al.}~\cite{TTO93} showed that  Sr$_{1-x}$La$_x$TiO$_3$
with perovskite-like structure exhibits a pronounced increase of the effective mass
in approaching the insulating state while keeping the existence of a large 
Fermi surface which satisfies the Luttinger sum rule. 
They already argued that the different behavior of this compound compared to the 
pronounced two dimensional high-T$_c$ cuprates might be due its   three dimensional 
character.
Furthermore, the    transfer of spectral weight in
 Li-doped NiO ~\cite{KKG89} and the investigation of hole and electron 
doping of this compound in Ref.~\onlinecite{RSH95}  
demonstrate the importance of strong electronic correlations in transition metal oxides.
A systematic analysis of the electronic structure of 3d- transition metal compounds was recently
performed by Saitoh {\em et al.}~\cite{SBM95}.
They showed that for a variety of materials like LaMnO$_3$,  LaFeO$_3$, LaCoO$_3$,
Mn$_{0.25}$TiS$_2$,  Ni$_{0.33}$TiS$_2$,   NiO, FeO, or MnO the character of the band gap
is of the charge transfer type.
Furthermore, the analysis of $x$-ray photoemission spectra~\cite{BMM96} demonstrates the
importance of  the charge transfer mechanism in systems like V$_2$O$_5$, TiO$_2$
or LaCrO$_3$ and showed that many of the early transition metal compounds
have to be reclassified as intermediate between the charge transfer and the Mott-Hubbard
region or as falling into the charge transfer region itself.
Consequently, it is  of importance for an understanding of these materials
 that one  takes the local spin  and charge fluctuations of the transition metal {\em and} the 
oxygen states  into account, i.e.  both orbital degrees of freedom have to be described
on the same footing.

For systems with one orbital degree of freedom a theoretical approach
that allows the  calculation of the excitation spectrum of strongly correlated  systems,
maintaining the dominating local  correlations, was recently proposed by 
Metzner and Vollhard~\cite{MV89}.
They   introduced a dynamical 
mean field theory, where the system can be mapped on 
a local problem coupled to an effective 
bath.~\cite{BM89,O91,GK92,J92} 
This dynamical mean field approach is exact in the limit of large 
coordination numbers or equivalently  for large spatial
dimensions ($D \rightarrow \infty$).~\cite{MV89,MH89}
In most applications of this approach, the one band Hubbard model is 
considered. A scaling of the hopping element 
like $t=t^*/\sqrt{D}$ with fixed $t^*$ leads for
$D \rightarrow \infty$ to a remarkable simplification of the 
many body problem while
retaining the nontrivial local dynamics of the 
elementary excitations and other main
features of the model.~\cite{GK92,J92,MH89,RZK92,GKr92,JP93,PCJ93} 

Only few attempts have been made to extend the dynamical mean field
theory to systems with more than one orbital degree of freedom.
Valenti and Gros~\cite{VG93} realized that for the two band Hubbard model with 
perovskite structure, a scaling procedure similar 
to the one band model does not lead to a finite bandwidth.  
In order to avoid these difficulties Georges {\it et al.}~\cite{GKK93}
introduced a CuO model where two interpenetrating hypercubic lattices 
of copper and oxygen sites have been considered.
This   leads to a finite bandwidth and is able to describe interesting
physics of the model under consideration.
However,  square root divergencies of the uncorrelated density of states occur,
which result from the sublattice structure of the underlying
lattice.
This could limit the range of applicability of this model  in the description
of real transition metal oxides or possibly high-T$_c$ systems.

In this paper we propose a scaling procedure to develop a reasonable model 
of perovskite systems for large $D$ based on a selective treatment of different
hopping processes in large dimensions.
It will be shown that this procedure leads to a large $D$ version of the three-band Hubbard
model which reproduces important physics of the low dimensional situation.
First, we solve this model using   perturbation theory~\cite{GK92,YY70}  and show
that this leads to a qualitatively wrong description of the physics near half filling.
These problems are avoided by using 
  an extended version of the non crossing 
approximation~\cite{KK71,MH84,BCW87,B87,PG89}.
Finally, we discuss the doping dependence of the
transition metal and oxygen densities of states.
This is of interest due to the evolution of coherent quasiparticles near the Fermi
energy, the transfer of spectral weight and the change of the effective mass upon doping.
\section{Theory}
\label{sec:Theory}
The dynamical mean field theory, which becomes exact in the 
limit of large coordination numbers, turned out to be a natural
 starting point for the description of highly correlated electronic
 systems like the one band Hubbard model and the Anderson lattice model.
All this is based on a proper scaling of the hopping element, 
originally proposed in Ref.~\onlinecite{MV89}.
In the following, we show that the extension of this approach 
to more complex systems is only possible if one generalizes 
the idea of the scaling procedure in a way that the corresponding 
transfer functions between different  unit cells, 
but not the hopping element itself, have to be scaled.
This is demonstrated for the case of the two band Hubbard 
Hamiltonian, which describes hybridized transition metal (TM)
and oxygen states with perovskite lattice structure.
Similar to the one band case, the problem can be mapped 
onto an Anderson impurity problem and the enormous amount 
of theoretical tools, developed  for the solution of this model, can be used.
Furthermore, we show that the most natural choice of the 
corresponding impurity problem is to
include all local orbitals into the impurity, not solely the 
strongly correlated ones.
Otherwise, the set of self consistent equations is  not well defined,
leading to various computational problems.
In our case, a cluster of hybridized TM- and O-sites is 
embedded into an effective medium.
Physically, this is necessary because the charge 
and spin fluctuations of these two states are closely 
intertwined and have to be described on the same footing.
The solution of the extended Anderson impurity model 
is performed using a perturbational approach~\cite{GK92,YY70}
 and an extended version of the
 non crossing approximation~\cite{KK71,MH84,BCW87,B87,PG89}.
 The    details of the  latter method are presented in the appendix. 
\subsection{Scaling procedure for large $D$}
We consider the  $D$-dimensional extension of the perovskite lattice, where 
the transition metal  
sites  sit on a $D$-dimensional hypercubic lattice and the $D$ oxygen sites per unit cell
are located between every two nearest neighbor TM sites. For simplicity, we consider 
only one TM 3d orbital.
The corresponding ($D+1$)-band Hubbard Hamiltonian for the 
TM 3d and O 2p$_{x_\alpha}$  ($\alpha=1, \ldots , D$) orbitals  reads
\begin{eqnarray}
H &=& \sum_{i,\sigma}\,\varepsilon_d   
      d_{i\sigma}^\dagger d_{i\sigma}\,
    + \,\sum_{l,\sigma, \alpha}\,\varepsilon_p    
      p_{l\sigma x_\alpha}^\dagger p_{l\sigma x_\alpha}\, \nonumber \\
    & &+ \,t\sum_{il,\sigma,\alpha}\, g^\alpha_{il} 
      (d_{i\sigma}^\dagger p_{l\sigma x_\alpha}\, + H.c.)
%\nonumber\\
%  & & 
+\,U \sum_i  n_{i \uparrow}^d n_{i \downarrow}^d  \, .
\label {Hubb}
\end{eqnarray}
\noindent Here,     $d_{i\sigma}^\dagger$ 
($p_{l\sigma x_\alpha}^\dagger$) creates 
a hole in a TM-3d (O-2p$_{x_\alpha}$ ) orbital at  site $i$ ($l$) with spin 
$\sigma$. $\varepsilon_d$ and $\varepsilon_p$ are the corresponding 
on-site energies. 
$t$ is the amplitude of the nearest neighbor TM-O hopping integral and $U$
is the Coulomb repulsion between TM holes with occupation number operator
$n_{i  \sigma}^d=d_{i\sigma}^\dagger d_{i\sigma}$.
The $p_{x_\alpha}-d$  hopping matrix elements have $p d \sigma$-symmetry, so that
$g^\alpha_{i l}=+1$  or $-1$ if $l=i+x_\alpha/2$ or $l=i-x_\alpha/2$, respectively.
Only the combination
%\begin{equation}
$p_{{\rm s} i \sigma}= \frac{1}{\sqrt{2 D}} \sum_{\l \alpha} \, 
g^\alpha_{i l} p_{l \sigma x_\alpha}$
%\end{equation}
of oxygen states hybridizes with the d-states~\cite{ZR88}.
Since the operators $p_{{\rm s} i \sigma}$ do not anticommute, it is necessary to 
orthogonalize
the corresponding states leading to 
$p_{{\bf k} \sigma}=\gamma_{\bf k}^{-1} p_{{\rm s} {\bf k} \sigma}$
where   $p_{{\rm s} {\bf k} \sigma}$ is the Fourier transform of $p_{{\rm s} i \sigma}$ and 
\begin{equation}
\gamma_{\bf k}^2=1 - \frac{1}{D} \sum_{\alpha=1}^D \cos k_\alpha\, .
\end{equation}
Besides the $p_{{\bf k} \sigma}$, which now fulfill    standard fermion commutation 
relations,   $D-1$ linear independent combinations of 
 the Fourier transforms of $ p_{l\sigma x_\alpha}^\dagger$
  occur, which are orthogonal to 
 $p^\dagger_{{\bf k} \sigma }$ and build up  nonbonding oxygen orbitals.
These nonbonding states at $\varepsilon_p$  decouple from the remainder 
of the system and do not have to be considered in the following calculation.
The resulting Hamiltonian reads:
\begin{eqnarray}
H &=& \sum_{{\bf k}\sigma} \left( \varepsilon_d 
d^\dagger_{{\bf k} \sigma} d_{{\bf k} \sigma} 
+ \varepsilon_p p^\dagger_{{\bf k} \sigma} p_{{\bf k} \sigma} 
 \right) 
\nonumber \\
 & & + \sqrt{2 D} t \sum_{{\bf k}\sigma}
 \left( \gamma_{\bf k} d^\dagger_{{\bf k} \sigma} p_{{\bf k} \sigma} 
 + H. c. \right) \, 
%\nonumber \\
%& & 
%+\,U \sum_i n_{i \uparrow}^d n_{i \downarrow}^d  \, .
\label{Htwob}
\end{eqnarray}
This is a straightforward extension of the two band Hubbard model of
 perovskite systems  to arbitrary 
dimension and can be used as the basis of a $1/D$ expansion.
However, there exists no scaling of the form $t=t^\ast/D^\beta $ 
with exponent $\beta$
 leading to a nontrivial limit for $D \rightarrow \infty$.
Valenti and Gros~\cite{VG93} suggested a scaling with $\beta=1/2$ which 
leads to a zero  bandwidth  of the bonding and
antibonding states, i.e. the TM-O "dimers" of different lattice sites decouple.
All other choices of   $\beta $ lead either to zero or to infinite density of 
states.

\begin{figure}
\centerline{\epsfig{file=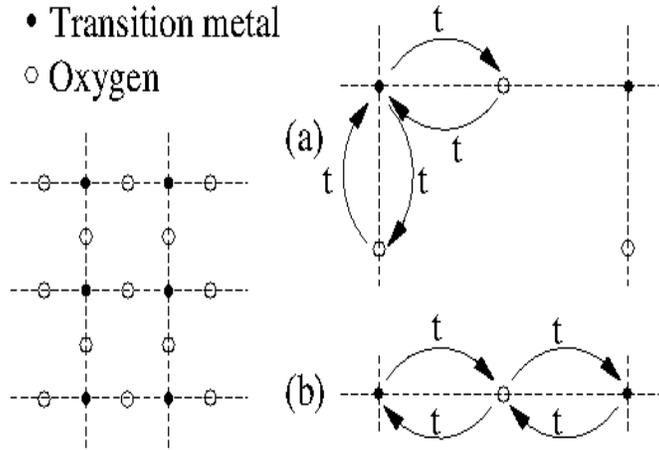,width=9cm,height=6cm}}
%\hspace{2.5cm}\epsfxsize=10cm \epsfbox{hopping.eps}
\caption{Hopping processes of order $t^4$ which are of different importance for large $D$.
Since there are $4D^2$ processes of type (a) but only $2D$ processes of type (b), the 
latter are suppressed compared to the former for large dimensions.}
\label{fig1}
\end{figure}

In order to obtain a physically reasonable limit of  the perovskites 
for large $D$ one has to bear in mind
that different hopping processes which are of  the same order in $t$  are of 
different importance for large $D$. 
Due to the local character of the Coulomb interaction, it is sufficient
to perform the scaling at first for $U=0$.
For the calculation of the single particle Green's function one has to sum
up all closed paths on the lattice under consideration.
In Fig. \ref{fig1} we show two different hopping processes which are of forth order in $t$.
There are $4D^2$ processes of  type (a) and only $2D$ processes
of type (b). Using  a scaling like $t=t^\ast/\sqrt{D}$ the contribution of the (a) processes to the
Green's function are of order $1$ whereas
the (b) processes  have only contributions   of order $1/D$ and are suppressed for $D \rightarrow \infty$ leading to the 
zero bandwidth of Ref.~\onlinecite{VG93}.
This results from the  TM coordination number being
 $2D$ whereas the oxygen coordination number is $2$.
A selective consideration of these different paths can be obtained by the following partial summation
of paths starting at TM-site $i$ and ending at the TM-site $j$:
\begin{eqnarray}
 & &G^o_{d\,ij}(\omega)= G^{at}_{d}(\omega) \delta_{ij} + G^{at}_{d}(\omega)T_{ij}(\omega) 
G^{at}_{d}(\omega) \nonumber \\
     \lefteqn{   +\sum_l\, G^{at}_{d}(\omega)T_{il} (\omega)G^{at}_{d}(\omega)
         T_{lj}(\omega) G^{at}_{d} (\omega) \,\,\, +\cdots  \,\,\, .}
\label{gdij}
\end{eqnarray}
Here, the local d-Green's function
\begin{equation}
G^{at}_{d}(\omega) =\left(\omega -\varepsilon_{d} -\frac{2Dt^2}
{\omega-\varepsilon_p}\right)^{-1}
\end{equation}
contains all TM-O-TM hopping processes, where the path returns to the 
starting TM-site without
entering any other TM site ((a) processes).
The transfer function $T_{ij}(\omega)$ which is given by
 $(-t^2)/(\omega-\varepsilon_p)$ 
 for neighboring TM sites  and zero otherwise has the character of
 an effective, but frequency dependent 
TM-TM hopping element $T_{ij}  \propto \delta_{\langle ij\rangle}  t^2$ ((b) processes).
Therefore, we propose the following scaling procedure, where the two different 
hopping processes are treated differently:
\begin{equation}
G^{at}_{d}(\omega) =\left( \omega -\varepsilon_{d} -\frac{4(t^*)^2}
{\omega-\varepsilon_p}\right)^{-1} \, \, ; \,\,\,\,\,\,\,  t^*=
\sqrt{D/2} t \, ,
\label{scal1a}
\end{equation}
\begin{equation}
T_{ij}(\omega)=-\delta_{\langle ij\rangle} \sqrt{\frac{2}{D}} 
\frac{(t^*)^2}{\omega - \varepsilon_p}
\,\, ; \,\,\,\,\,\,\,\,\, (t^*)^2=\sqrt{D/2} t^2 \, .
\label{scal1b}
\end{equation}
Here, the effective TM-TM hopping  $T_{ij}$ (not $t$ itself) has been scaled
 in analogy to the one band case.
Consequently, the processes of Fig. \ref{fig1}b are no more suppressed relative to that of 
Fig. \ref{fig1}a.
The summation of all closed paths on the remaining hypercubic TM-lattice
is straightforward and leads to the d-Green's function for $U=0$:
\begin{equation}
G^o_d(\omega)=\frac{\omega-\varepsilon_p}{4 (t^*)^2} \int dy \frac{\varrho_o(y-1)}
{(\omega-\varepsilon_p)(\omega-\varepsilon_d)/4 (t^*)^2 -y} \, ,
\label{GFo}
\end{equation}
where the reduced density of states $\varrho_o(y)$ is defined by
\begin{equation}
\varrho_o(y)= \frac{1}{N} \sum_{\bf k} \, 
\delta \left(y-\frac{1}{\sqrt{2D}} \alpha_{\bf k}\right)\, ,
\label{reddos}
\end{equation}
with $\alpha_{\bf k}= \sum_{\nu=1}^{D} {\rm cos}k_\nu $.
The oxygen Green's function $G^o_p(\omega)$ which refers to the states 
$p_{{\bf k} \sigma}$
can be obtained from  $G^o_d(\omega)$ by interchanging $\varepsilon_d$ and 
 $\varepsilon_p$.
The large $D$ behavior of $\varrho_o(y)$ is, in analogy to the one band case,
given by a Gaussian distribution function 
$\sqrt{2/\pi} \exp(-2y^2)$.~\cite{MH89}
For simplicity, we use in the following a semi elliptical  reduced 
density of states $\varrho_o(y)=2/\pi \sqrt{1-y^2}$ if $y^2 <1$ and zero
 otherwise.
This Bethe lattice of the TM-sites  does not changes the relative local
 arrangement of TM and O sites, typical for 
perovskite systems.
Finally, it is interesting to point out, that this selected treatment of
different hopping processes can be generated by the Hamiltonian of 
Eq.~\ref{Htwob}, if one performs the following scaling transformation of 
the coherence factor:
\begin{equation}
t^2 D \gamma_{\bf k}^2 \rightarrow  2(t^*)^2 - 
2\sqrt{\frac{2}{D}}(t^*)^2 \alpha_{\bf k}\, .
\end{equation}

Based on this formulation of the large $D$ version of the TM-O system 
for $U=0$, it is straightforward to  show for $U \neq 0$, 
in analogy to the one band case, that the self energy $\Sigma_d(\omega)$ 
of TM-holes is momentum independent.
Consequently, one finds for the TM-Green's function within the Bethe lattice:
\begin{eqnarray}
G_d(\omega)^{-1}& &=\omega-\varepsilon_d-\Sigma_d(\omega)-\frac{4t^2}{\omega-\varepsilon_p}
\nonumber \\ & &
-\frac{1}{4} \left(\frac{4t^2}{\omega-\varepsilon_p}\right)^2 G_d(\omega) .
\label{semi_ell_rel}
\end{eqnarray}
This results from the representation of the Green's function with semi-elliptical density of states:
\begin{eqnarray}
G_d(\omega) &=& \frac{1}{N}\sum_{\bf k} G_d({\bf k},\omega) \nonumber \\
& &= \frac{\omega-\varepsilon_p}{4t^2} {\cal F}\left( 
\frac{(\omega-\varepsilon_p)(\omega-\varepsilon_d-\Sigma_d(\omega))}{4t^2} \right) ,
\end{eqnarray}
with ${\cal F}(z)\equiv 2\left( z-1-\sqrt{(z-1)^2-1} \right)$ and the relation ${\cal F}(z)^{-1} = z-1-\frac{1}{4}{\cal F}(z)$.
Furthermore, the system can be mapped 
onto an Anderson model with effective hybridization
and additional self consistency condition.

Finally, we  note that the additional consideration of the  TM-O Coulomb 
repulsion $U_{pd}$ would lead
  to further {\em frequency dependent} contributions of the self energies 
$\Sigma_{ \alpha \beta}(\omega)$  ($ \alpha \beta \in \{p,d\}$).
This  results from the small coordination number of the oxygen sites and 
is different in the one band case, where each intersite Coulomb repulsion 
contributes in zeroth
order of $1/D$ only through its Hartree-Fock value~\cite{MH89}.
 Therefore, the limit $D \rightarrow \infty$ might be an interesting 
 approach for a detailed consideration of the dynamical interplay of spin and
charge fluctuations of transition metal oxides.  
 
\subsection{Effective TM-O impurity model}

\begin{figure}
\centerline{\epsfig{file=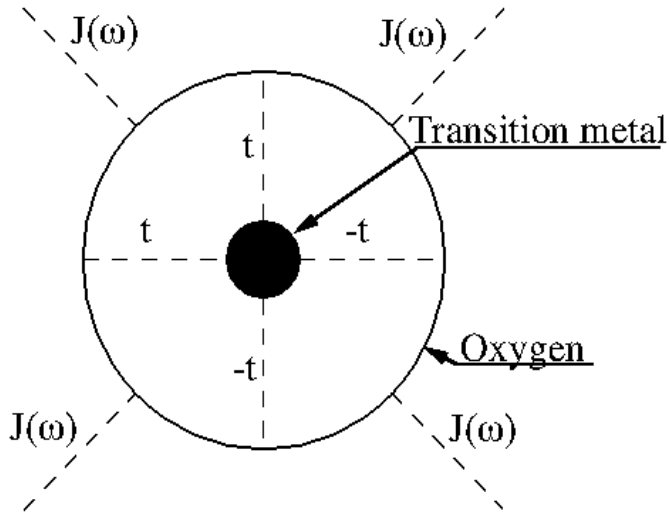,width=9cm,height=7cm}}
%\hspace{2.5cm}\epsfxsize=10cm \epsfbox{nca_hybrid.eps}
\caption{Schematic impurity model where a cluster of one 
TM- and O-site is embedded within an effective medium.  
Here, the local hybridization $t$ between 
the two local sites is explicitly taken into account. 
The effective medium is coupled only to the oxygen states.}
\label{fig2}
\end{figure}

As in the one band case, the large-D version of a TM-O model is mapped 
onto an effective impurity model. Due to the occurrence of an additional 
orbital degree of freedom, the choice of the corresponding effective medium 
is not unique. Although, the most obvious choice  might be to embed only the
TM-site into an effective medium and to change solely the self consistence
condition, it is straightforward to show that this leads 
to a divergence of the hybridization of this medium at the oxygen on-site 
energy $\varepsilon_p$.
This is due to the fact that the effective medium has to simulate not solely 
the correlated TM-sites but also the existence of the oxygen states. 
In order to avoid this divergency, we propose in the following an  impurity 
model where a cluster of one TM- and one O-orbital is embedded within 
an effective medium.
This is illustrated in Fig.~\ref{fig2}. Motivated by the perovskite geometry, 
we couple the effective medium only to the oxygen sites. Furthermore, 
the local TM-O hybridization within the cluster is explicitly taken into account.
This leads to a well defined effective medium. 
The corresponding Hamiltonian reads:
\begin{eqnarray}
H_{\rm eff}=H_{\rm loc}+H_{\rm med} \, ,
\label{Imp_Hamil}
\end{eqnarray}
where the local part is given by
\begin{eqnarray}
H_{\rm loc} =  \sum_\sigma \left( \varepsilon_d n_\sigma^d 
+ \varepsilon_p n_\sigma^p
+ 2t\left( d_\sigma^\dagger p_\sigma + 
p_\sigma^\dagger d_\sigma \right) \right) 
+ U n_\uparrow^d n_\downarrow^d \nonumber
\end{eqnarray}
and
\begin{eqnarray}
H_{\rm med} = \sum_{{\bf{k}}\sigma} \left( W_{\bf{k}} c_{{\bf{k}}\sigma}^\dagger p_\sigma  
+ H.c. \right)
+ \sum_{{\bf{k}}\sigma} \varepsilon_{\bf{k}} c_{{\bf{k}}\sigma}^\dagger c_{{\bf{k}}\sigma}
\nonumber
\end{eqnarray}
describes the coupling of the oxygen states with the effective medium,
 characterized by the hybridization
\begin{eqnarray}
{\cal J}(\omega)=\sum_{\bf{k}} \frac {|W_{\bf{k}}|^2}{\omega+i0^+-\varepsilon_{\bf{k}}} \, .
\end{eqnarray}
This coupling of the effective medium to the oxygen states, although not unique,
is physically motivated by the local arrangement of the original lattice.
From the equation of motion of the effective Hamiltonian, it follows
\begin{eqnarray}
G_d(\omega)^{-1}=\omega-\varepsilon_d-\Sigma_d(\omega)-\frac{4t^2}
{\omega-\varepsilon_p-{\cal J}(\omega)} \, .
\label{Eq_motion}
\end{eqnarray}
Comparing this with Eq.~\ref{semi_ell_rel}, one immediately finds:
\begin{eqnarray}
{\cal J}(\omega)= t^2 G_d(\omega) \left( 1-\frac{t^2 G_d(\omega)}
{\omega-\varepsilon_p-t^2 G_d(\omega)} \right) \, .
\label{self_cons}
\end{eqnarray}
This is the self-consistant equation of the theory, based on the condition that the Green's function of 
the impurity model defined in Eq.~\ref{Imp_Hamil}  equals that of the lattice system.
 Note that this effective 
medium, which  represents the states coupled to the oxygen, is  predominantly 
determined by the  correlated TM-sites.

In the following we solve the impurity model of Eq.~\ref{Imp_Hamil} using the iterated 
perturbation theory (IPT), and an extended version of the non crossing approximation 
(NCA), where the effective medium is fixed by Eq.~\ref{self_cons}.
Since it was necessary for the consideration of the TM-O cluster to generalize the 
standard NCA-scheme to more local eigenstates,
 the corresponding details are given in the appendix.

\section{Results}
\begin{figure}
%\hspace{2.5cm}\epsfxsize=10cm \epsfbox{u=8ipt.eps}
\centerline{\epsfig{file=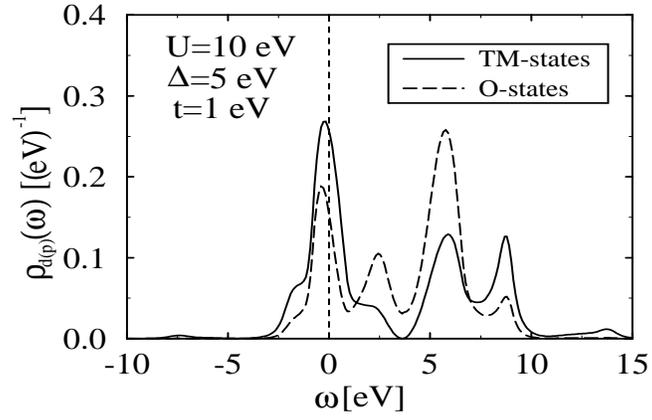,width=9cm,height=7cm}}
\caption{TM (solid line) and O (dashed line) density of states obtained
within the iterated perturbation theory for half filling.
Note that the spectral weight of the various bands is erroneously described
within the IPT such that no insulator occurs for half filling.}
\label{fig3}
\end{figure}

In the following, we present our   results for the TM- and O-densities of states
obtained within    the IPT and NCA.  
In Fig. ~\ref{fig3} we show our results for the TM- and O-densities of states 
obtained within the IPT for half filling $x=0$ ($x=n_d+n_p-1$).
Two Hubbards bands which are dominated by TM-states are separated by $U$ and two
oxygen dominated bands in the neighborhood of $\varepsilon_p$ are visible.
Similar to the one band case, the IPT leads to an interesting structure  of 
the density of states, where now four bands are occurring.
However, the expected insulating behavior at half filling (for large $U$) 
does not occur.
This results from the overestimation of the spectral weight of the lower 
Hubbard band within the IPT by $\propto 1-n_d$. 
The occupation of oxygen sites ($n_p >0$) due to  TM-O hybridization 
leads to an overcounting of copper sites which can be occupied without 
paying any Coulomb energy.
Therefore, the  success of the IPT in the one band case~\cite{GK92}
 does not occur within the two band model. This is due to the absence 
of the particle-hole symmetry  and the change of the TM-occupation 
number $n_d$ as function of $t$  at half filling. This shortcoming of 
the IPT results from the occupation of O sites and
is expected to vanish for small $n_p$ which occurs in the limit of a
 large value for $\Delta=\varepsilon_p-\varepsilon_d$.
However, for the physically interesting situation \mbox{$\Delta \approx U/2$},
the IPT leads to qualitatively wrong results and one has to develop
theoretical approaches  which take the local TM-O many body states 
explicitly into account. 
Therefore, we  use the NCA , which was shown to be in excellent 
agreement with quantum Monte Carlo (QMC) simulations in the 
one band case, such that a local
TM-O hybrid  with  16 local eigenstates is coupled to an effective 
medium. Here, the TM-O singlet  state which  is  related to an important
excitation near half filling  is explicitly taken into account.

\begin{figure}
\hskip  3.45cm
\epsfig{file=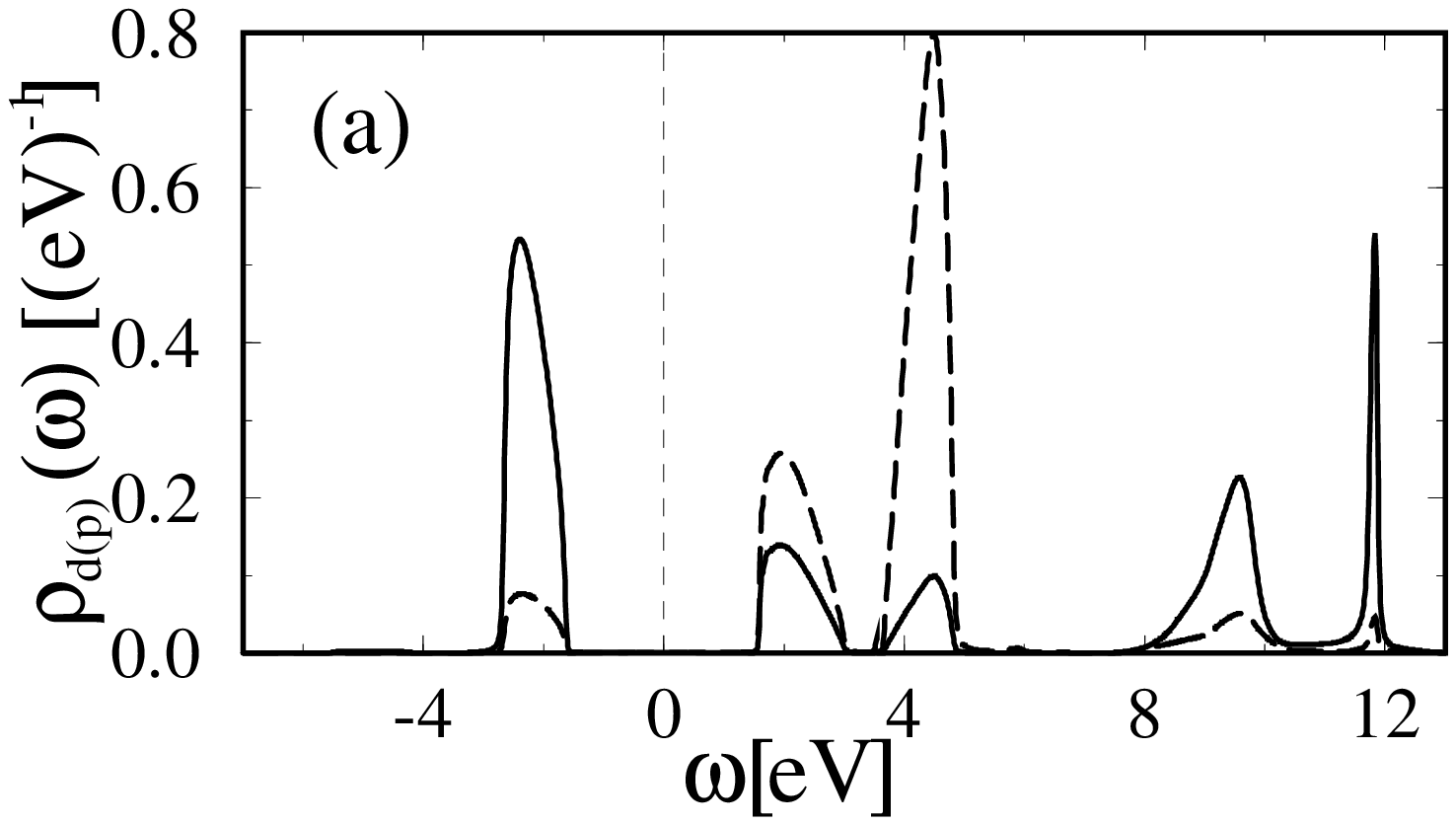,width=9cm,height=17cm}
%\hskip +1.5cm
\vskip -1.5cm
\centerline{\epsfig{file=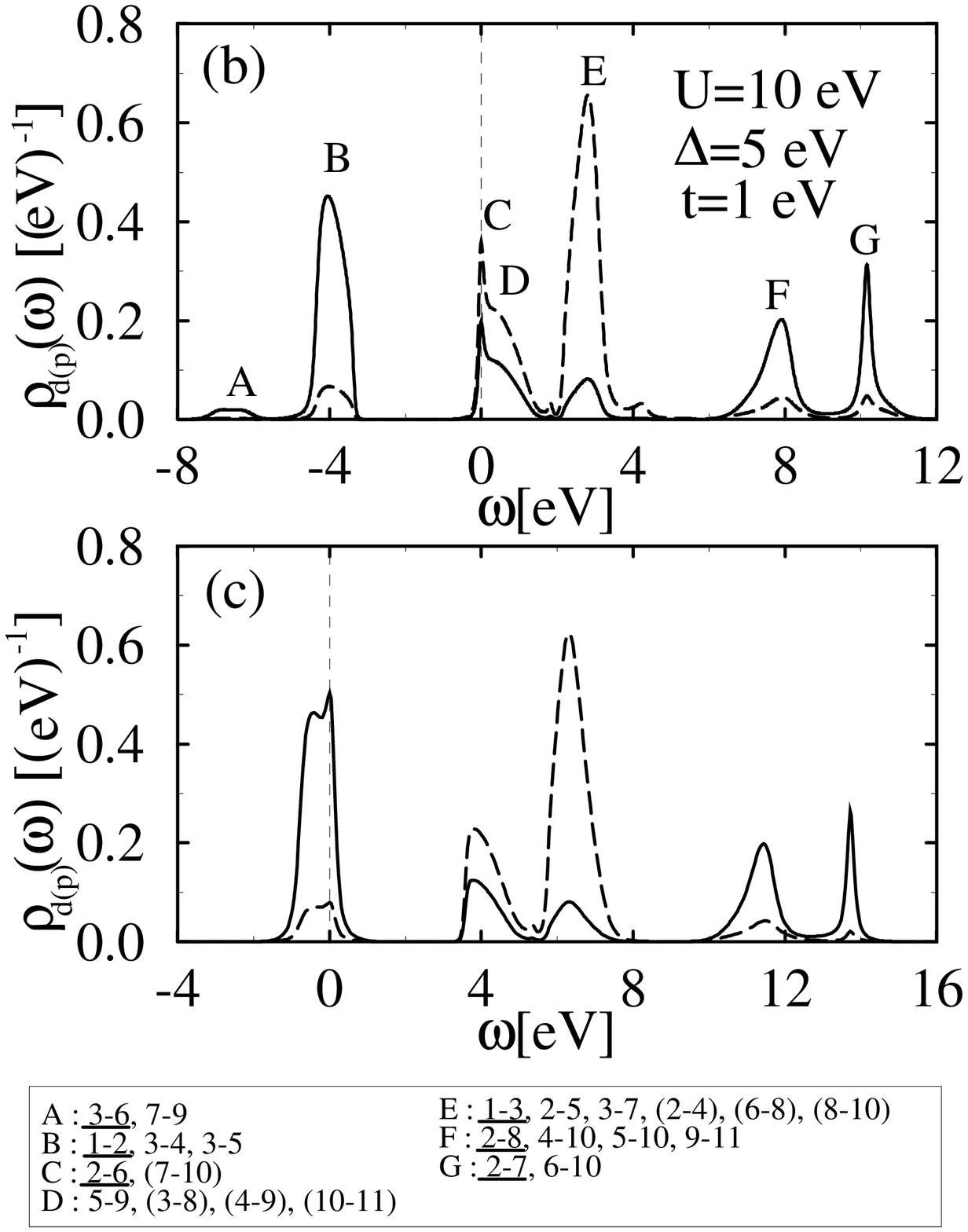,width=9cm,height=15cm}}
\vskip -1cm 
\caption{TM (solid line) and O (dashed line) density of states obtained
within the hybrid-NCA at half filling (a) for hole doping $x=0.1$ (b) and 
electron doping $x=-0.1$ (c).
  For hole doping it is indicated which transitions between the local states contribute
 to the corresponding bands, where the meaning of the state labels is given in Tab. 1.
The leading contribution is underlined while negligible contributions are embraced.}
\label{fig4}
\end{figure}

 In Fig. ~\ref{fig4}  we show the TM- and O-densities of states at half filling and
for a hole and an electron doped system, obtained
 within  the   non crossing approximation.
The calculations are performed for $U=2 \Delta =10 t$ and   $t=1\, {\rm eV}$.
At half filling we find now the expected insulating state, i.e. the Fermi level lies in the 
middle of the charge transfer gap.
In this charge transfer regime we find bands dominated by TM-states (B, F, and G)
which are separated
by the Coulomb repulsion $U$ due to the Mott-Hubbard splitting and on both
sides of the charge transfer gap  TM- and O-dominated bands (B and C).
Besides the O-dominated band above the charge transfer gap (C-D) a second one for higher
excitation energy occurs (E). A detailed analysis of the transitions between the various
many body states of the local TM-O hybrid reveals that these are build up by 
the TM-O singlet and triplet states.
All these features of the density of states are believed to be general properties of the
model under consideration~\cite{BSB93} and are also  qualitatively in agreement with exact
cluster diagonalizations  and Monte Carlo simulations for two dimensional  
systems.~\cite{WHS91,TM92,DMH92}
This demonstrates that it is indeed possible to reproduce important physical
phenomena of the low dimensional situation within the large $D$ approach.
Besides this general structure on a larger energy scale, a coherent quasi particle
peak near the Fermi level occurs, which shows up an interesting temperature and
doping dependency (see below).

In order to clarify the   origin of the details of the density of states,
we  investigated which transitions between the local many body states are
 predominately
responsible for each band. 
Technically, this is performed within the NCA if one restricts the available Hilbert space
according to the general scheme presented in the appendix. If the spectral weight of a  
band decreases after projecting out a certain local state $| m \rangle$, a transition
from or into this state is responsible for the band.
This analysis is of particular importance, if one is interested  in calculations
within a restricted Hilbert space of the most important states.
The results are presented in Fig.~\ref{fig4}b for hole doped systems, whereby the meaning of the 
state labels are  given in   Tab.~\ref{tab1}.
The leading contribution is underlined while negligible contributions are embraced.

The main contributions to the bands B and E occur also for an uncorrelated system and
are build up by  transitions between  an empty state $| 0 \rangle $ and  a bonding
 ($| d_\sigma \rangle $)  or antibonding ($| p_\sigma \rangle $) combination 
of  TM- and O-states, respectively.
More interesting are the bands A, C, F, and G, because they are purely due to   strong
electronic correlations.
Here, the bands F and G are due to transitions from the TM-dominated state 
$| d_\sigma \rangle $ to states which are dominated by doubly occupied TM sites, i.e.
they form the upper Hubbard band.
Band C  contains essentially the well known TM-O Zhang Rice singlet state~\cite{ZR88}
 resulting from a transition 
from the bonding state $| d_\sigma \rangle $ to the state $|S_o \rangle $ dominated
by  the TM-O singlet state.
 other contributions close in energy are identified under label D.
Furthermore, a pronounced part of band E, which consists predominantly of antibonding
$| p_\sigma \rangle $ states, results also from  the transition of singly occupied sites to
TM-O triplet states $|T_\pm \rangle $ .
Consequently, we find a singlet triplet splitting which compares well with the result
$8 J$ obtained from perturbation theory in $t$ for two spatial dimensions, where 
$J=t^2(\Delta^{-1} + (U-\Delta)^{-1})$.~\cite{ZO88}
Finally, the lowest state A results from a transition between the antibonding
state $| p_\sigma \rangle $ and the singlet state $| S_o \rangle $.
Interestingly, this state occurs for very low excitation energies and only 
in   systems with an occupied singlet state, i.e. for hole doped systems.
This is a consequence of the fact that the singlet state with two particles per TM-O unit
has lower energy than the singly occupied antibonding state $| p_\sigma \rangle $.

\begin{figure}
\centerline{\epsfig{file=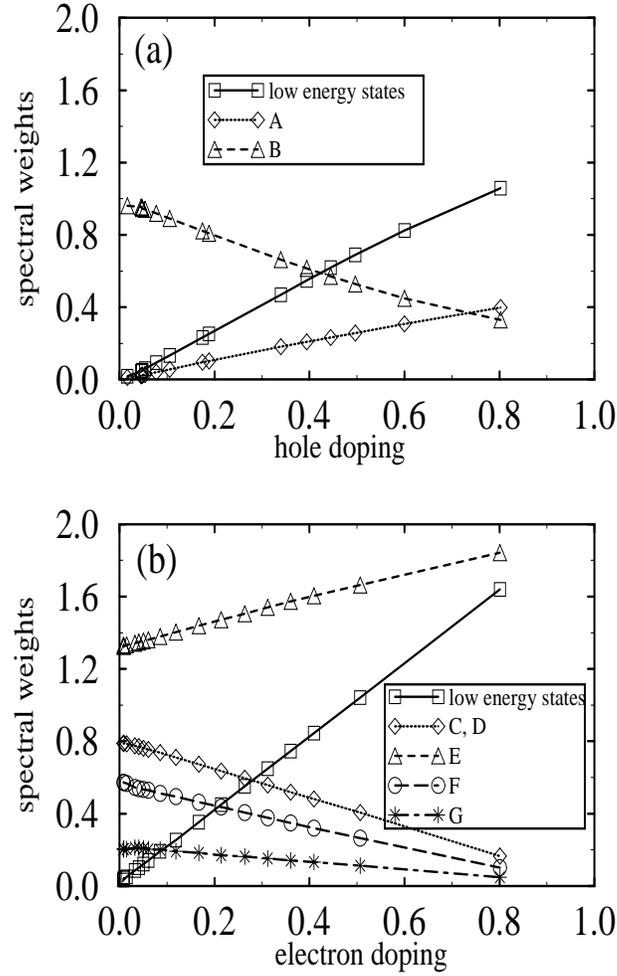,width=9cm,height=15cm}}
\caption{Doping dependence of the spectral weight multiplied
with the spin degeneration of the occupied bands and 
the occupied part of the singlet band  (low energy hole states) for hole doping (a) 
and of the unoccupied  bands and the unoccupied  part of the lower Hubbard band 
(low energy electron  states) for electron doping (b).}
\label{fig5}
\end{figure}

This interpretation of the various bands in terms of the transitions between many body
states allows a transparent explanation of the doping dependence of their spectral weight.
In  Fig.~\ref{fig5} we show the spectral weight of various bands upon electron and hole doping.
For hole doping we plot in part a of the figure the spectral weight of all occupied states 
and of the occupied part   of the singlet  band (low energy hole states).
The latter demonstrates that the transfer of spectral weight, discussed in Ref.~\onlinecite{EMS91},
is    included within our approach.
As  expected, the spectral weight of the bonding band decreases    upon doping.
However, band A which  results from the transition  between $| S_o \rangle $ and 
$| p_\sigma \rangle $ increases, because upon hole doping more initial singlet states 
 are available
The corresponding result for the electron doped materials are shown in   Fig.~\ref{fig5}b,
where we plot the spectral weight of all unoccupied bands and that of the unoccupied part of
the lower Hubbard band B which carries the Fermi level (low energy electron states).
While the spectral weight of the bands which are purely due to correlations decreases,
band E increases upon electron doping.
The character of the system becomes much more uncorrelated leading to an increase
of the bonding and antibonding TM-O states which are present also for $U=0$ with even
larger spectral weight.

\begin{figure}
\centerline{\epsfig{file=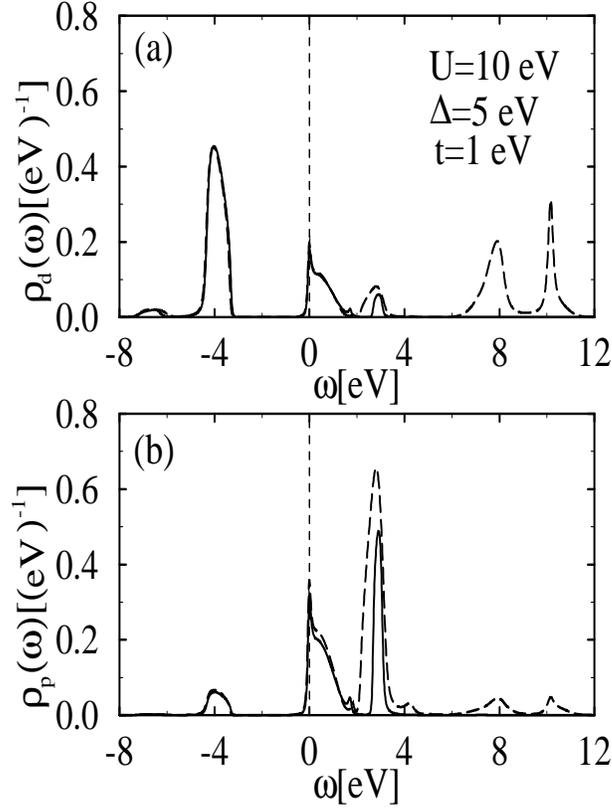,width=9cm,height=15cm}}
\vskip -3cm
 \caption{ TM (a) and O (b) density of states obtained
within the hybrid-NCA for hole doping $x=0.1$ within a restricted Hilbert space of
the local states $|0 \rangle$, $|d_\sigma \rangle$, $|p_\sigma \rangle$, $|T_\sigma \rangle$, 
and $|S_o \rangle$ in comparison with the results of the full local space (dashed lines).}
\label{fig6}
\end{figure}

Based on this analysis, which demonstrates that the density of states near the Fermi level is
build up by the transitions between only   a few local many body states, it is interesting to
perform the calculation within a restricted Hilbert space.
In Fig.~\ref{fig6}  we show the TM- and O-densities of states if one takes only the states
$| 0 \rangle$, $| p_\sigma \rangle$,$| d_\sigma \rangle$,$| S_o \rangle$, and
$| T_\pm \rangle$ into account.
The results are compared with that  for the full Hilbert space (dashed line).
It is impressive, that the details of the single particle excitation spectrum  near the Fermi level
are very well reproduced by this restricted set of   states.
Note that this calculation includes still the full information about the orbital
character of the low energy spectrum.
This might be of interest for the theoretical description of more complex systems
where the diagonalization of the local problem leads to extremely large numbers of
coupled integral equations of the NCA.
The  most important states can straightforwardly be found from the eigenstates of the 
complete local problem, since the peak position is well reproduced from the corresponding
difference of the relative eigenenergies, as given in Tab.~\ref{tab1}.

\begin{figure}
\centerline{\epsfig{file=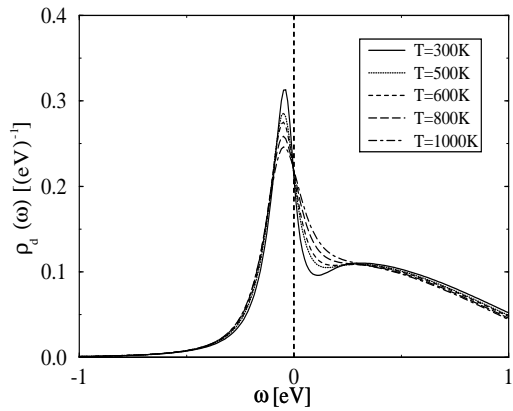,width=9cm,height=7cm}}
\caption{Density of states near the Fermi level for hole doping $x=0.2$ and 
for various temperatures.
Note the formation of a coherent quasi particle peak resulting from Kondo like resonances
 between the singly occupied bonding TM-O state and the doubly occupied singlet state. }
\label{fig7}
\end{figure}

Finally, we present our results for the coherent low energy peak near the Fermi level.
In Fig.~\ref{fig7} we show the density of states near the 
Fermi level ($\omega=0$) for a hole doping concentration $x=0.2$  
for various temperatures~\cite{NOTE_NCA}.
For electron doping, the  corresponding low energy part of the spectrum behaves very
similarly.
Although the density of states at the Fermi level is almost constant, the states near the 
Fermi energy are strongly temperature dependent.
For decreasing temperature, a coherent quasi particle peak occurs which builds up
its maximum slightly below the Fermi energy.

\begin{figure}
\centerline{\epsfig{file=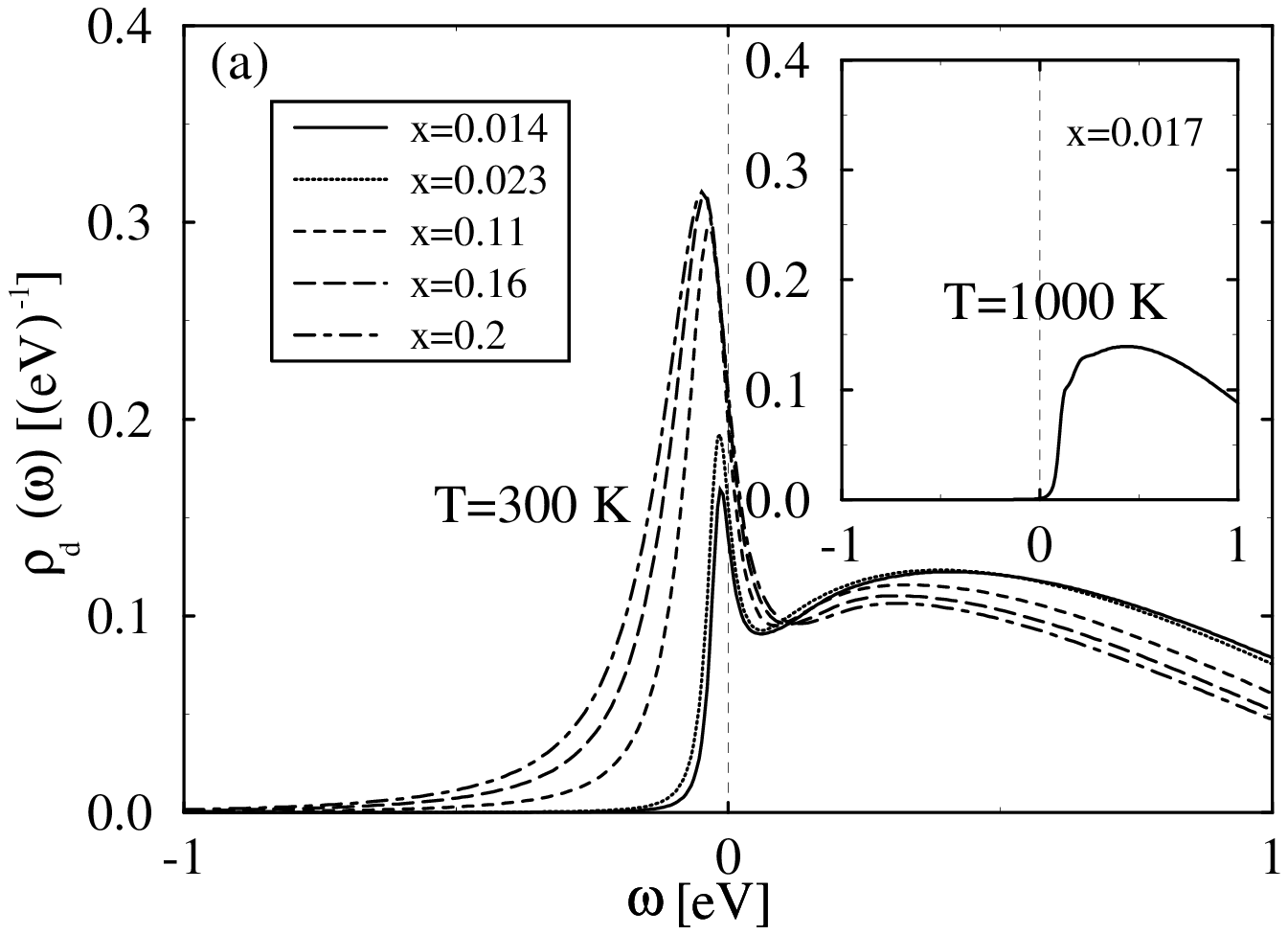,width=9cm,height=7cm}}
\vskip -2cm
\hskip 3.55cm
\epsfig{file=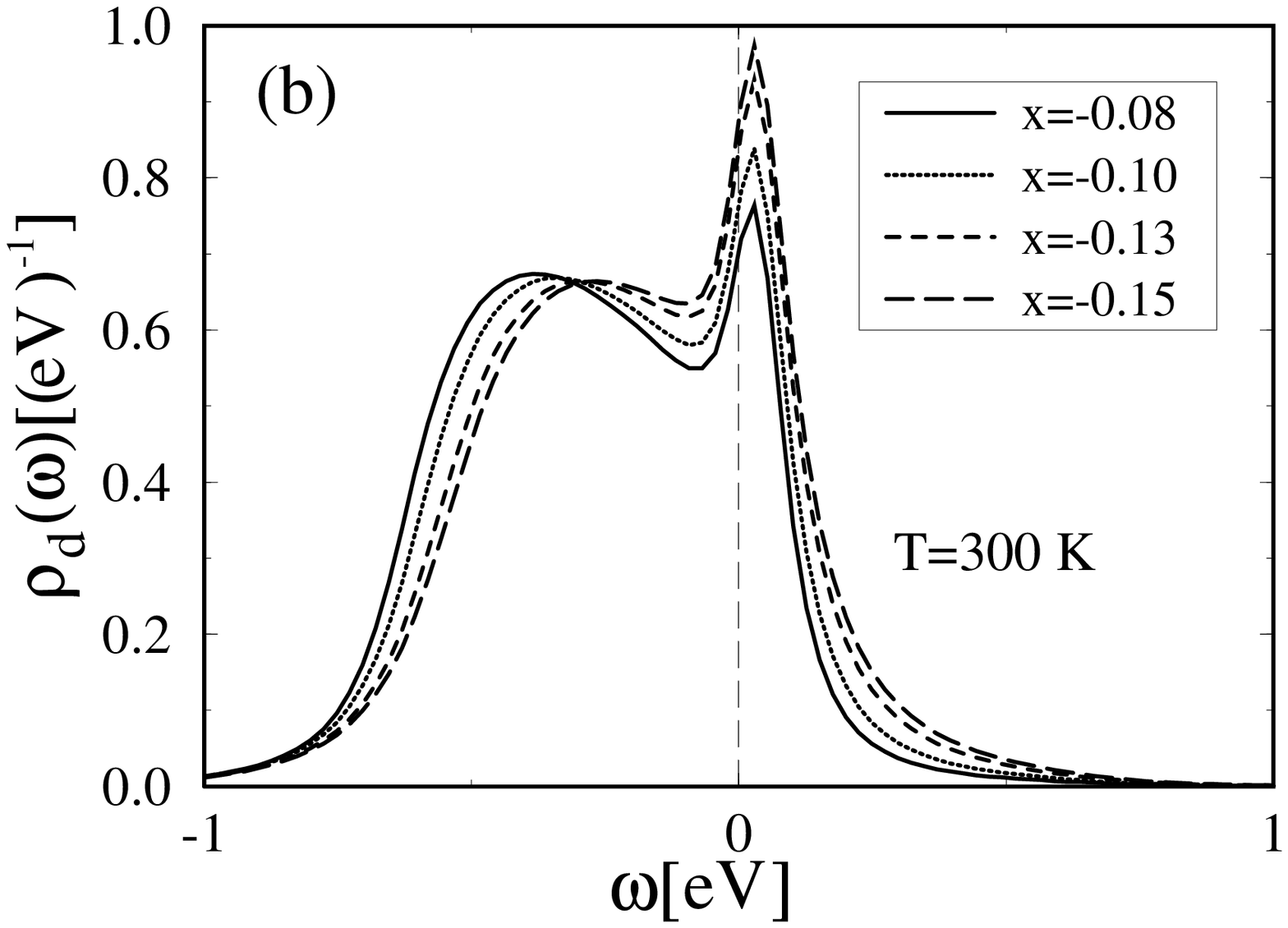,width=8cm,height=11cm}
\vskip -5cm
\hskip 3.55cm
\epsfig{file=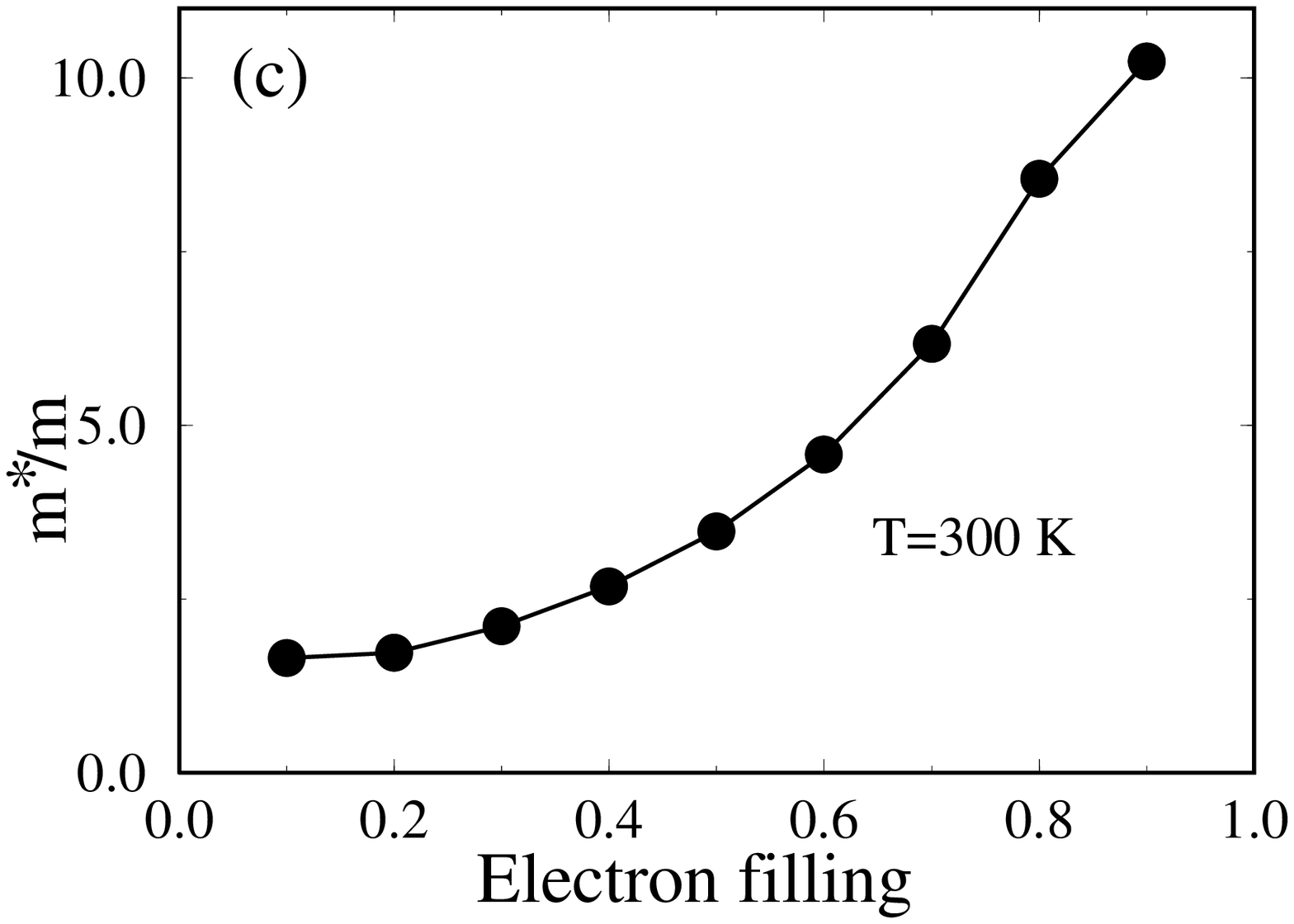,width=8cm,height=11cm}
\vskip -2cm
\caption{Doping dependence of the coherent quasi particle peak for $T=300\, {\rm K}$
in the charge transfer regime with $U=10 \, {\rm eV}$,  $ \Delta=5  \, {\rm eV}$ (a)
and in the Mott-Hubbard regime with $U=5 \, {\rm eV}$,  $ \Delta=9.5  \, {\rm eV}$ (b).
In both cases we use $t=1\, {\rm eV}$.
The effective mass in the charge transfer regime and for  electron doping is shown in part (c).
For comparison we also  plot in the inset the result for low hole doping 
($x=0.017$) but high temperature  $T=1000\, {\rm K}$ which demonstrates that 
strong temperature dependence 
occurs in the low doping regime. }
\label{fig8}
\end{figure}

In Fig.~\ref{fig8}a and b we present our results for the doping dependence of the low energy part of 
the spectrum for a hole doped system in the charge transfer and for
an electron doped system in the Mott-Hubbard regime at $T=300\, {\rm K}$. 
 As expected, the general behavior of the low energy part of the total density of states 
is relatively  similar in both regimes.
The evolution of these quasiparticles results in a large effective mass ratio obtained
via $m^*/m=\left. (1-\partial {\rm Re}\Sigma_d(\omega)/\partial \omega) \right|_{\omega=0}$.
In Fig.~\ref{fig8}c,
as a representive example we show  results for a electron doped material in the 
charge transfer regime. 
This shows clearly the mass enhancement when approaching the insulating regime either
charge transfer or Mott Hubbard type and resembles the doping dependence of  the effective mass 
obtained within 
the Brinkmann-Rice transition~\cite{BR70,BAR89}, 
where only the coherent part of the spectrum occurs.
In addition, we plot  in part a of   Fig.~\ref{fig8}   the density of states for 
low hole doping ($x=0.017$) but for large
temperatures of $T=1000\, {\rm K}$.
This shows that the quasi particle peak disappears completely between $300 $
 and  $1000\, {\rm K}$, giving rise to an enormous temperature dependence of 
 the low energy excitations in slightly doped perovskite systems.
The enhancement of $m^*/m$ with band filling agrees with the observed behavior
  in   Sr$_{1-x}$La$_x$TiO$_3$~\cite{TTO93}
which is believed to fall into the the Mott-Hubbard regime.
Furthermore, the measurement of the Hall coefficient indicates the formation of a 
large Fermi surface.
Although our dynamical mean field approach does not allow an investigation of the 
momentum dependent spectral density 
\[
A({\bf k},\omega)\approx \frac{m}{m^*} \delta(\omega-\varepsilon^*_{\bf k}) +
A^{\rm incoh.}({\bf k},\omega) \, ,
\]
we believe that a  large effective mass ratio $m^*/m $ leads to a large Fermi 
surface  formed by the coherent part of the spectrum 
$\frac{m}{m^*} \delta(\omega-\varepsilon^*_{\bf k})$ with renormalized 
quasi particle energy $\varepsilon^*_{\bf k}$.
This is possible because only  a few particles can 
occupy a large number of  momentum states due to its low spectral weight 
$ \frac{m}{m^*}$.
 The physical origin of the coherent quasiparticle resonance  is   similar
to   the conventional Abrikosov-Suhl resonance of systems with  magnetic 
impurities.
In the charge transfer regime, 
for hole doping, the resonant transition between a bonding state $| d_\sigma \rangle$ and
 the Zhang Rice singlet   $| S_o \rangle$ builds up the Kondo like peak.
For electron doping, it is correspondingly a transition between an empty state $|0 \rangle$
and $| d_\sigma \rangle$.

\section{Conclusions}
In conclusion, we presented a   dynamical mean field theory for perovskites which is based
on a  new scaling procedure of the large dimensional Hubbard model  of transition metal
3d and oxygen 2p states.
In the limit $D \rightarrow \infty $, the electronic self energy is, similar to the one band case,
momentum independent and the model can be mapped onto an  Anderson impurity model.
It is demonstrated that the most natural choice of this impurity is to include all
local orbitals of one unit cell, not solely the strongly correlated ones.
Correspondingly a model with   16 local many body states is embedded into an effective 
medium which has to be  determined self consistently.
Motivated by the perovskite lattice geometry, this effective medium is coupled only to
the oxygen sites of the impurity and simulates predominantly the TM-sites of the real 
lattice.
The solution of the local problem is performed using the iterated perturbation theory and
an extended    noncrossing approximation.
Is is shown that the spectral weight of the various bands of the density of 
states is erroneously described within the IPT.
This shortcoming results from the missing electron hole symmetry of the half filled model,
leading to an absence of a metal insulator transition for half filling.
In distinction, the NCA is able to describe  this problem properly and gives interesting insight
into the origin and doping dependence of the various bands of the TM- and O-densities of
states.
Furthermore, the occurrence  and doping as well as temperature dependencies of a coherent
quasi particle peak near the Fermi level is discussed.
This results from a Kondo like resonance between an empty and a bonding TM-O state for
electron doping and this bonding state and the Zhang Rice singlet state for hole doping.
Finally, it is shown that the restriction of the local states to a limited set of most important
states reproduces very well the low energy density of states, keeping the full orbital information
of the involved states.
This is of importance for an extension of the theory to more realistic situations including all 
five d-bands such that only a small number of coupled NCA equations have to be solved.
This might be stimulating for further theoretical studies of transition metal compounds 
within the dynamical mean field approach.

\acknowledgments
Part of the work of J. S. has been done at LEPES-CNRS, Grenoble. 
Support by the Commission of the European Communities under
 EEC-Contract CHRX-CT93-0332 is gratefully acknowledged.

%\newpage
\appendix

\section{Non Crossing Approximation of a TM-O Cluster}

In this appendix, we discuss an extended version of the non crossing 
approximation for an impurity model given in Eq.~\ref{Imp_Hamil}.
The NCA was shown to be in excellent agreement with quantum 
Monte-Carlo calculations for the dynamical mean-field theory of the one band 
case~\cite{JP93,PCJ93}.
This approximation is not a perturbational expansion with respect 
to the Coulomb repulsion U, but takes into account all eigenstates 
of the local part of the impurity Hamiltonian, by performing an
 expansion with respect to the hybridization ${\cal J}(\omega)$.
Therefore, we rewrite the Hamiltonian of Eq.~\ref{Imp_Hamil} in 
terms of the Hubbard operators $X_{m' m}\equiv |m'\rangle\langle m|$,
 where $|m\rangle$ is, in our case, one of the 16 eigenstates of $H_{loc}$, 
with eigenvalue $E_m$ (see Tab.~\ref{tab1}):
\begin{eqnarray*}
H_{loc} &=& \sum_{\sigma m} E_m  X_{m m} \, ,\\
H_{med} &=& \sum_{{\bf{k}}\sigma m m'} \left( 
W_{\bf{k}} c_{{\bf{k}}\sigma} ^\dagger 
P_{m m'}^\sigma X_{m m'}  + H.c.\right) +
\sum_{{\bf{k}}\sigma} \varepsilon_{\bf{k}} c^\dagger_{{\bf{k}}\sigma} c_{{\bf{k}}\sigma} \, .
\end{eqnarray*}
The eigenvalues of $H_{loc}$ are explicitly given in Ref.~\onlinecite{N91}.
The  $(16 \times 16)$ matrix  $P^\sigma$ and a corresponding matrix $D^\sigma$
are defined by the following representation of the   O- and  TM-destruction 
operators:
\begin{eqnarray*}
p_\sigma&=&\sum_{m,m'} P_{m m'}^\sigma X_{m m'} \, ,\\
d_\sigma&=&\sum_{m,m'} D_{m m'}^\sigma X_{m m'} \, .
\end{eqnarray*}
Using this representation of the Hamiltonian, one introduces within 
NCA scheme the local propagators for each of the eigenstates $|m\rangle$:
\begin{eqnarray}
P_m(\omega)=\frac{1}{\omega+i0^+-E_m-\Sigma_m(\omega)} \, ,
\label{NCAprop}
\end{eqnarray}
with the corresponding spectral density 
$ p_m(\omega)=-\frac{1}{\pi}~{\rm Im} P_m(\omega+i0^+) $. 
The NCA self energies are given by:
\begin{eqnarray}
\Sigma_m(\omega)&=& -\frac{1}{\pi} \sum_{\sigma , m'} |P_{m,m'}^\sigma|^2 \int_{-\infty}^{+\infty}
d\varepsilon f\left( \eta_{m,m'} \varepsilon \right) \nonumber \\
& &{\rm Im} {\cal J}(\varepsilon+i0^+)  
 P_{m'}\left( \omega+\eta_{m,m'} \varepsilon \right)\, .
\label{NCAself}
\end{eqnarray}
Here, $f(\varepsilon)$ is the Fermi distribution function and 
$\eta_{m,m'}=+1$ (resp. $-1$) if the particle number of the 
state $|m'\rangle$ is higher (resp. lower) than the particle 
number of the state $|m\rangle$.
In Eq.~\ref{NCAself}, we neglect the vertex corrections 
discussed in  Refs.~\cite{PG89,KQ90,AG94,A95}.
Although this was shown to slightly underestimate the 
Kondo temperature and to shift the Abrikosov-Suhl resonance 
slightly away from the chemical potential, it gives important 
insights into the low energy excitations, in particular for large 
values of the Coulomb repulsion.
Furthermore, in view of the enormous numerical problems, 
which arizes if one calculates these vertex corrections even for less complex systems than 
ours,~\cite{AG94,A95}  and  due to the additional self consistent condition
of the large $D$ approach, it is actually very difficult to go beyond 
Eq.~\ref{NCAself}.
Finally, the Green's functions of the system, which give the 
information about the single particle excitation spectrum, result from:
\begin{eqnarray}
G_d(\omega)&=&\sum_{m,m'} |D_{m m'}^\sigma|^2 \langle 
\langle  X_{m' m} ; X_{m m'} \rangle\rangle_\omega \, ,\\
G_p(\omega)&=&\sum_{m,m'} |P_{m m'}^\sigma|^2 \langle 
\langle  X_{m' m} ; X_{m m'} \rangle\rangle_\omega \, ,
\end{eqnarray}
where $\langle \langle  X_{m' m} ; X_{m m'} 
\rangle\rangle_\omega$ is the Fourier transform of the 
retarded Green's function $-i \theta (t) \langle [ X_{m' m}(t) ;
 X_{m m'}(0) ]_+ \rangle$. They can be expressed with respect 
to the NCA local propagators as following:

\begin{eqnarray}
\langle \langle  X_{m' m} ; X_{m m'} \rangle\rangle_\omega & =&
 \frac{1}{Z_0}\int_{-\infty}^{\infty} d\varepsilon
{e}^{-\beta \varepsilon} \left( p_{m'}(\varepsilon) P_{m}(\omega+\varepsilon)
\right. \nonumber \\ 
& &  \left. - p_{m}(\varepsilon) P_{m'}(\omega-\varepsilon)^* \right) \, ,
\end{eqnarray}
where the partition sum $Z_0$ is defined by 
$\sum_m\int_{-\infty}^{\infty} d\varepsilon {e}^{-\beta \varepsilon} 
p_{m}(\varepsilon)$. All the summations with respect to $m$ and 
$m'$ are performed over the 16 eigenstates of $H_{loc}$. 
As can be seen from Tab.~\ref{tab1} the  spin degeneracy leads to only 11 
different eigenstates and propagators   in the paramagnetic state.
The solution of Eq.~\ref{NCAprop} and \ref{NCAself},
which are coupled singular integral equations, is performed using
the fast Fourier transformation for the convolution integrals and 
using the defect propagators of the local eigenstates~\cite{MH84,BCW87}.

\newpage  

\newpage
\end{multicols}
 
\begin{table}
\begin{tabular}{|c||l|c|c|c|}               
\hline
state label &  state &   degeneracy  & occupation  $n_o$ & $E_m-\mu n_o $ \\
 \hline  \hline
1    &   $| 0 \rangle   $    & 1  & 0 & 0 \\
2    &   $| d_\sigma \rangle=(\cos\phi d^\dagger_\sigma+\sin\phi p^\dagger_\sigma) | 0 \rangle $
                       & 2  & 1 & -0.701 \\
3    &   $| p_\sigma  \rangle =(\cos\phi d^\dagger_\sigma-\sin\phi p^\dagger_\sigma) | 0 \rangle $ 
                       & 2  & 1 & 5.701 \\
4    &   $| T_o \rangle=\frac{1}{\sqrt{2}}( d^\dagger_\uparrow p^\dagger_\downarrow 
                      +
                d^\dagger_\downarrow p^\dagger_\uparrow) | 0 \rangle $ 
                       & 1 & 2 & 5 \\
5    &  $| T_ \sigma \rangle=d^\dagger_\sigma p^\dagger_\sigma | 0 \rangle $   
                      & 2  & 2 & 5 \\
6    &   $| S_o \rangle =[ u_1  d^\dagger_\uparrow d^\dagger_\downarrow
                                          + v_1 p^\dagger_\uparrow p^\dagger_\downarrow
                                     $  & 1  & 2 & 2.738 \\  & 
                   $  \ \ \ \  \ \ \   - w_1 ( d^\dagger_\uparrow p^\dagger_\downarrow -
                  d^\dagger_\downarrow p^\dagger_\uparrow)]  | 0 \rangle $
                     && &   \\  
7    &   $| dd \rangle = [ u_2  d^\dagger_\uparrow d^\dagger_\downarrow
                                           + v_2 p^\dagger_\uparrow p^\dagger_\downarrow
                                         $  & 1  & 2 & 12.217 \\  &    
                                        $  \ \ \ \ \ \ \  - w_2 ( d^\dagger_\uparrow p^\dagger_\downarrow -
                                        d^\dagger_\downarrow p^\dagger_\uparrow)]  | 0 \rangle $
                       & & &  \\ 
8    &   $|  pp  \rangle = [ u_3  d^\dagger_\uparrow d^\dagger_\downarrow
                                          + v_3 p^\dagger_\uparrow p^\dagger_\downarrow
                                  $  & 1  & 2 & 10 \\ & $   \ \ \ \   \ \ \   - w_3 ( d^\dagger_\uparrow
                                               p^\dagger_\downarrow -
                                        d^\dagger_\downarrow p^\dagger_\uparrow)]  | 0 \rangle $ 
                & & &  \\ 
9    &   $| D_\sigma \rangle =(  \cos \theta  d^\dagger_\sigma p^\dagger_\sigma 
 d^\dagger_{- \sigma} 
$ & & &  \\ & $ \ \ \ \ \ \ \ 
+ \sin \theta  d^\dagger_\sigma p^\dagger_\sigma 
 p^\dagger_{- \sigma} ) | 0 \rangle  $
               & 2  & 3 & 9.298 \\
10    &   $| P_\sigma  \rangle= (  \cos \theta  d^\dagger_\sigma p^\dagger_\sigma 
 p^\dagger_{- \sigma}$ & & &  \\ & $ \ \ \ \ \ \ \  - \sin \theta  d^\dagger_\sigma p^\dagger_\sigma 
 d^\dagger_{- \sigma} ) | 0 \rangle$ 
              & 2 & 3 & 15.702 \\
11    &   $| DP \rangle=d^\dagger_\uparrow p^\dagger_\uparrow d^\dagger_\downarrow 
p^\dagger_\downarrow    | 0 \rangle $ 
              & 1  & 4 & 20 \\
\hline
\end{tabular}
\caption{The degeneracy, occupation and relative energy of all local eigenstates
 introduced in the appendix  and the corresponding labels used in Fig. 4 are given.
The eigenvalues are obtained for $U=10\, {\rm eV}$, $\Delta=5\, {\rm eV}$, and
$t=1\, {\rm eV}$.
Here, $\tan 2\phi=-\frac{4 t}{\Delta}$ and $\tan 2\theta=\frac{4 t}{\Delta+U}$ .
The coefficients $u_i$, $v_i$, and $w_i$ can  be obtained from the diagonalization
of the three dimensional space build up by $ | S_o \rangle $,  $ | dd \rangle $, and
 $ | pp \rangle $, see  for example Ref. \protect\onlinecite{N91}.}
\label{tab1}
\end{table}
\newpage 
\vskip 2cm
\newpage
$ \ \ \ $
\vskip 4cm
\newpage
$ \ \ \ $
\vskip -10cm

\newpage
 
\newpage


\begin{references}
\bibitem{NOTE_NCA} We chose $T=300\, {\rm K}$ as  lowest temperature 
                                         such that our results are believed to be not affected
                                         by the low temperature power law singularity of the 
                                         density of states of the NCA, see Ref.~\onlinecite{MH84}.
\end{references}
\end{document}